\documentclass[preprint,3p,times]{elsarticle}

\usepackage{amssymb}
\usepackage{amsmath}
\usepackage[all,cmtip]{xy}
\usepackage{amsthm}
\usepackage{color}
\usepackage{enumitem}
\usepackage{tikz}
\usepackage[colorlinks=true]{hyperref}
\usepackage[hyphenbreaks]{breakurl}
\input diagxy

\theoremstyle{plain}
  \newtheorem{thm}{Theorem}[section]
  
  \newtheorem{prop}[thm]{Proposition}
  
\theoremstyle{definition}

  \newtheorem{rem}[thm]{Remark}

\DeclareMathAlphabet{\mathcal}{OMS}{cmsy}{m}{n}

\DeclareMathOperator{\id}{id}

\makeatletter
\def\ps@pprintTitle{%
 \let\@oddhead\@empty
  \let\@evenhead\@empty
  \def\@oddfoot{\vbox{\hsize=\textwidth\footnotesize
  \vskip 8pt
  \copyright 2020. This manuscript version is made available under the CC-BY-NC-ND 4.0 license \url{https://creativecommons.org/licenses/by-nc-nd/4.0/}. The published version is available at \url{https://doi.org/10.1016/j.fss.2020.03.007}.\\
  }}%
  \let\@evenfoot\@oddfoot}
\makeatother

\def\oto{{\bfig\morphism<180,0>[\mkern-4mu`\mkern-4mu;]\place(86,0)[\circ]\efig}}

\newcommand{\da}{\downarrow}
\newcommand{\ua}{\uparrow}

\newcommand{\lra}{\longrightarrow}
\newcommand{\lda}{\swarrow}
\newcommand{\lua}{\nwarrow}

\newcommand{\bs}{\backslash}

\newcommand{\bv}{\bigvee}
\newcommand{\bw}{\bigwedge}

\newcommand{\dv}{\dashv}

\newcommand{\nat}{\natural}

\renewcommand{\phi}{\varphi}

\newcommand{\lam}{\lambda}

\newcommand{\ka}{\kappa}

\newcommand{\CD}{\mathcal{D}}

\newcommand{\CL}{\mathcal{L}}

\newcommand{\CQ}{\mathcal{Q}}

\newcommand{\sK}{{\sf K}}

\newcommand{\sM}{{\sf M}}

\newcommand{\sP}{{\sf P}}

\newcommand{\sy}{{\sf y}}

\newcommand{\Fix}{{\sf Fix}}

\newcommand{\FQ}{\mathfrak{Q}}

\newcommand{\Cat}{{\bf Cat}}

\newcommand{\Dist}{{\bf Dist}}

\newcommand{\Rel}{{\bf Rel}}

\newcommand{\QCat}{\CQ\text{-}\Cat}

\newcommand{\QDist}{\CQ\text{-}\Dist}

\newcommand{\QRel}{\CQ\text{-}\Rel}

\newcommand{\dphi}{\phi^{\da}}
\newcommand{\uphi}{\phi_{\ua}}

\newcommand{\sPd}{\sP^{\dag}}
\newcommand{\syd}{\sy^{\dag}}

\newcommand{\PX}{\sP X}
\newcommand{\PY}{\sP Y}

\newcommand{\PdX}{\sPd X}
\newcommand{\PdY}{\sPd Y}

\newcommand{\Mphi}{\sM\phi}

\newcommand{\Kphi}{\sK\phi}
\newcommand{\Kdphi}{\sK^{\dag}\phi}

\newcommand{\ldd}{\mathrel{/}}
\newcommand{\rdd}{\mathrel{\bs}}
\newcommand{\with}{\mathrel{\&}}

\newcommand{\Qw}{\CQ_{\with}}

\newcommand{\QFL}{\CQ^F_{\CL}}
\newcommand{\QPL}{\CQ^P_{\CL}}
\newcommand{\QOL}{\CQ^O_{\CL}}

\newcommand{\ro}{{\rm ob}}
\newcommand{\arr}{{\rm arr}}

\numberwithin{equation}{section}

\begin{document}

\begin{frontmatter}



\title{Multi-adjoint concept lattices via quantaloid-enriched categories}


\author{Hongliang Lai}
\ead{hllai@scu.edu.cn}

\author{Lili Shen\corref{cor}}
\ead{shenlili@scu.edu.cn}

\cortext[cor]{Corresponding author.}
\address{School of Mathematics, Sichuan University, Chengdu 610064, China}

\begin{abstract}
With quantaloids carefully constructed from multi-adjoint frames, it is shown that multi-adjoint concept lattices, multi-adjoint property-oriented concept lattices and multi-adjoint object-oriented concept lattices are derivable from Isbell adjunctions, Kan adjunctions and dual Kan adjunctions between quantaloid-enriched categories, respectively.
\end{abstract}


\end{frontmatter}




\section{Introduction}

The theory of quantaloid-enriched categories, initiated by Walters \cite{Walters1981}, established by Rosenthal \cite{Rosenthal1996} and developed by Stubbe \cite{Stubbe2005,Stubbe2006}, has revealed itself to be a fundamental and powerful toolkit in the study of fuzzy set theory, especially in the fields of many-valued sets and many-valued preorders (see, e.g., \cite{GutierrezGarcia2018,Hoehle2015,Hoehle2011a,Pu2012}). The survey paper \cite{Stubbe2014} is particularly recommended as an overview of quantaloid-enriched categories for the readership of fuzzy logicians and fuzzy set theorists.

Based on the fruitful results related to many-valued preorders, the theory of quantaloid-enriched categories has provided a general categorical framework \cite{GutierrezGarcia2018,Lai2017,Shen2014,Shen2013a} for the study of formal concept analysis (FCA) \cite{Davey2002,Ganter1999} and rough set theory (RST) \cite{Pawlak1982,Polkowski2002}. Explicitly, given a small quantaloid $\CQ$, a \emph{$\CQ$-distributor}
$$\phi:X\oto Y$$
between \emph{$\CQ$-categories} $X$ and $Y$ may be thought of as a multi-typed and multi-valued relation that is compatible with the $\CQ$-categorical structures on $X$ and $Y$, and it induces three pairs of \emph{adjoint $\CQ$-functors} between the (co)presheaf $\CQ$-categories of $X$ and $Y$:
\begin{enumerate}[label=(\arabic*)]
\item the \emph{Isbell adjunction} \cite{Shen2013a} $\uphi\dv\dphi:\PdY\to\PX$,
\item the \emph{Kan adjunction} \cite{Shen2013a} $\phi^*\dv\phi_*:\PX\to\PY$,
\item the \emph{dual Kan adjunction} \cite{Shen2014} $\phi_{\dag}\dv\phi^{\dag}:\PdX\to\PdY$,
\end{enumerate}
where we denote by $\PX$ and $\PdX$ the presheaf $\CQ$-category and the copresheaf $\CQ$-category of $X$, respectively. If we consider a $\CQ$-distributor $\phi:X\oto Y$ as a multi-typed and multi-valued \emph{context} in the sense of FCA and RST, then the \emph{complete $\CQ$-categories} of fixed points of the above adjunctions, denoted by
$$\Mphi:=\Fix(\dphi\uphi),\quad\Kphi:=\Fix(\phi_*\phi^*)\quad\text{and}\quad\Kdphi:=\Fix(\phi^{\dag}\phi_{\dag}),$$
may be viewed as ``concept lattices'' of the context $(X,Y,\phi)$; indeed, if we assume that the $\CQ$-categories $X$ and $Y$ consist of \emph{properties} (also \emph{attributes}) and \emph{objects}, respectively, then $\Mphi$, $\Kphi$ and $\Kdphi$ present the categorical version of the \emph{formal concept lattice}, the \emph{property-oriented concept lattice} and the \emph{object-oriented concept lattice} of $(X,Y,\phi)$, respectively. The recent work \cite{Lai2017} of Lai and Shen establishes a general framework for constructing various kinds of representation theorems of such ``concept lattices''. In particular:
\begin{enumerate}[label=(\arabic*)]
\item If $\CQ={\bf 2}$, the two-element Boolean algebra, and $\phi$ is a binary relation between (crisp) sets $X$ and $Y$, then $\Mphi$, $\Kphi$ and $\Kdphi$ reduce to the formal concept lattice \cite{Ganter1999}, the property-oriented concept lattice and the object-oriented concept lattice \cite{Yao2004a,Yao2004} of the (crisp) context $(X,Y,\phi)$ in the classical setting.
\item If $\CQ=\FQ$ is a \emph{unital quantale} \cite{Rosenthal1990} and $\phi$ is a fuzzy relation between (crisp) sets $X$ and $Y$ (i.e., $\phi$ is a map $X\times Y\to\FQ$), then $\Mphi$, $\Kphi$ and $\Kdphi$ are concept lattices of the \emph{fuzzy context} $(X,Y,\phi)$ of (crisp) sets $X$ and $Y$ \cite{Bvelohlavek2004,Lai2009,Shen2013}.
\item If $\CQ=\CD\FQ$ is the quantaloid of \emph{diagonals} (cf. \cite{Hoehle2011a,Pu2012,Stubbe2014}) of a unital quantale $\FQ$ and $\phi$ is a fuzzy relation between fuzzy sets $X$ and $Y$ (cf. \cite[Definition 2.3]{GutierrezGarcia2018}), then $\Mphi$, $\Kphi$ and $\Kdphi$ are concept lattices of the \emph{fuzzy context} $(X,Y,\phi)$ of fuzzy sets $X$ and $Y$ \cite{GutierrezGarcia2018,Shen2014,Shen2013b}.
\end{enumerate}

Since 2006, the theory of \emph{multi-adjoint concept lattices} was introduced by Medina, Ojeda-Aciego and Ruiz-Calvi{\~ n}o \cite{Medina2012,Medina2007,Medina2009,Medina2006} as a new machinery of FCA and RST unifying several approaches of fuzzy extensions of concept lattices, and it has been studied in a series of subsequent works (see, e.g., \cite{Cornejo2015,Cornejo2015a,Cornejo2018,Cornelis2014,Medina2012a,Medina2016}). As the basic notion of this theory, an \emph{adjoint triple} \cite{Medina2007,Medina2009,Medina2006} $(\with,\lda,\lua)$ with respect to posets $L_1$, $L_2$, $P$ satisfies
$$x\with y\leq z\iff x\leq z\lda y\iff y\leq z\lua x$$
for all $x\in L_1$, $y\in L_2$, $z\in P$, which is similar to the adjoint properties possessed by every quantaloid (see \eqref{quantaloid-comp-imp} below). It is then natural to ask whether it is possible to incorporate the theory of multi-adjoint concept lattices into the general framework of quantaloid-enriched categories, and the aim of this paper is to provide an affirmative answer to this question.

With the necessary background on quantaloids and quantaloid-enriched categories introduced in Section \ref{Quantaloids}, in Section \ref{Multi-adjoint-frames} we carefully exhibit how an adjoint triple gives rise to a quantaloid of three objects (Proposition \ref{Qw-def}), based on which we formulate quantaloids
$$\QFL,\ \QPL,\ \QOL$$
out of a multi-adjoint frame, a multi-adjoint property-oriented frame and a multi-adjoint object-oriented frame $\CL$, respectively, in Propositions \ref{QFL-def}, \ref{QPL-def} and \ref{QOL-def}. In each of the three cases, a context $(X,Y,\phi)$ of the respective frame $\CL$ is expressed as a $\QFL$-relation $\phi_F:X\oto Y$, a $\QPL$-relation $\phi_P:X\oto Y$ and a $\QOL$-relation $\phi_O:X\oto Y$, respectively, in Propositions \ref{context-QFL-relation}, \ref{context-QPL-relation} and \ref{context-QOL-relation}. Therefore, in Sections \ref{FCA-Isbell} and \ref{RST-Kan} we are able to apply the constructions of Isbell adjunctions, Kan adjunctions and dual Kan adjunctions to $\phi_F$, $\phi_P$ and $\phi_O$, respectively, and obtain the following main results of this paper:
\begin{enumerate}[label=(\arabic*)]
\item The multi-adjoint concept lattice \cite{Medina2009} of a context $(X,Y,\phi)$ of a multi-adjoint frame $\CL$ is given by a fibre of the complete $\QFL$-category $\Mphi_F$ (Theorem \ref{FCA-lattice}).
\item The multi-adjoint property-oriented concept lattice \cite{Medina2012} of a context $(X,Y,\phi)$ of a multi-adjoint property-oriented frame $\CL$ is given by a fibre of the complete $\QPL$-category $\Kphi_P$ (Theorem \ref{RST-lattice-P}).
\item The multi-adjoint object-oriented concept lattice \cite{Medina2012} of a context $(X,Y,\phi)$ of a multi-adjoint object-oriented frame $\CL$ is given by a fibre of the complete $\QOL$-category $\Kdphi_O$ (Theorem \ref{RST-lattice-O}).
\end{enumerate}
These results, once again, illustrate the thesis of Lawvere that \emph{fundamental structures are themselves categories} \cite{Lawvere1973}.

\section{Quantaloids and quantaloid-enriched categories} \label{Quantaloids}

For the convenience of the readers, in this section we recall the basic notions of quantaloids and quantaloid-enriched categories, and also fix the notations.

\subsection{Quantaloids}

A \emph{quantaloid} $\CQ$ \cite{Rosenthal1996,Stubbe2014} is a category whose hom-sets are complete lattices, such that the composition $\circ$ of $\CQ$-arrows preserves arbitrary joins on both sides, i.e.,
$$v\circ\Big(\bv_{i\in I} u_i\Big)=\bv_{i\in I}v\circ u_i\quad\text{and}\quad\Big(\bv_{i\in I} v_i\Big)\circ u=\bv_{i\in I}v_i\circ u$$
for all $u,u_i\in\CQ(p,q)$, $v,v_i\in\CQ(q,r)$ $(i\in I)$. Hence, the corresponding Galois connections induced by the compositions
$$\bfig
\morphism/@{->}@<4pt>/<800,0>[\CQ(q,r)`\CQ(p,r);-\circ u]
\morphism(800,0)|b|/@{->}@<4pt>/<-800,0>[\CQ(p,r)`\CQ(q,r);-\,\ldd\,u]
\place(400,5)[\mbox{\footnotesize{$\bot$}}]
\efig\quad\text{and}\quad\bfig
\morphism/@{->}@<4pt>/<800,0>[\CQ(p,q)`\CQ(p,r);v\circ -]
\morphism(800,0)|b|/@{->}@<4pt>/<-800,0>[\CQ(p,r)`\CQ(p,q);v\,\rdd\,-]
\place(400,5)[\mbox{\footnotesize{$\bot$}}]
\efig$$
satisfy
\begin{equation} \label{quantaloid-comp-imp}
v\circ u\leq w\iff v\leq w\ldd u\iff u\leq v\rdd w
\end{equation}
for all $u\in\CQ(p,q)$, $v\in\CQ(q,r)$, $w\in\CQ(p,r)$, where the operations $\ldd$ and $\rdd$ are called \emph{left} and \emph{right implications} in $\CQ$, respectively.

Let $\CQ_{\ro}$ denote the class of objects of a quantaloid $\CQ$. For each $p,q\in\CQ_{\ro}$, we denote by $\bot_{p,q}$ the bottom element of the hom-set $\CQ(p,q)$, and by $\id_q$ the identity $\CQ$-arrow on $q$. A quantaloid $\CQ$ is \emph{non-trivial} if
$$\bot_{q,q}<\id_q$$
for all $q\in\CQ_{\ro}$, since $\bot_{q,q}=\id_q$ would force every hom-set $\CQ(p,q)$ or $\CQ(q,r)$ $(p,r\in\CQ_{\ro})$ to contain only one element, i.e., $\bot_{p,q}$ or $\bot_{q,r}$.

\subsection{$\CQ$-relations}

From now on we let $\CQ$ denote a \emph{small} quantaloid $\CQ$; that is, $\CQ_{\ro}$ is assumed to be a \emph{set} instead of a proper class. In this case, the class $\CQ_{\arr}$ of $\CQ$-arrows of $\CQ$ is also a set.

Given a (``base'') set $T$, a set $X$ equipped with a map $|\text{-}|:X\lra T$ is called a \emph{$T$-typed set}, where the value $|x|\in T$ is the \emph{type} of $x\in X$, and we write
$$X_q:=\{x\in X\mid |x|=q\}$$
for the \emph{fibre} of $X$ over $q\in T$.

Considering $\CQ_{\ro}$ as the set of types, a \emph{$\CQ$-relation} (also \emph{$\CQ$-matrix} \cite{Heymans2010})
$$\phi:X\oto Y$$
between $\CQ_{\ro}$-typed sets $X$, $Y$ is a map
$$\phi:X\times Y\to\CQ_{\arr}\quad\text{with}\quad\phi(x,y)\in\CQ(|x|,|y|)$$
for all $x\in X$, $y\in Y$. With the pointwise local order
$$\phi\leq\phi':X\oto Y\iff\forall x,y\in X:\ \phi(x,y)\leq\phi'(x,y)\ \text{in}\ \CQ(|x|,|y|)$$
inherited from $\CQ$, the category $\QRel$ of $\CQ_{\ro}$-typed sets and $\CQ$-relations becomes a (large) quantaloid in which
\begin{align}
&\psi\circ\phi:X\oto Z,\quad(\psi\circ\phi)(x,z)=\bv_{y\in Y}\psi(y,z)\circ\phi(x,y),\label{QRel-comp}\\
&\xi\ldd\phi:Y\oto Z,\quad(\xi\ldd\phi)(y,z)=\bw_{x\in X}\xi(x,z)\ldd\phi(x,y),\label{QRel-left-imp}\\
&\psi\rdd\xi:X\oto Y,\quad(\psi\rdd\xi)(x,y)=\bw_{z\in Z}\psi(y,z)\rdd\xi(x,z) \label{QRel-right-imp}
\end{align}
for all $\CQ$-relations $\phi:X\oto Y$, $\psi:Y\oto Z$, $\xi:X\oto Z$, and
$$\ka_X:X\oto X,\quad\ka_X(x,y)=\begin{cases}
\id_{|x|}, & \text{if}\ x=y,\\
\bot_{|x|,|y|}, & \text{else}
\end{cases}$$
serves as the identity $\CQ$-relation on $X$.

\begin{rem}
$\CQ$-relations between $\CQ_{\ro}$-typed sets may be thought of as \emph{multi-typed} and \emph{multi-valued} relations. Indeed, a $\CQ$-relation $\phi:X\oto Y$ may be decomposed into a family of $\CQ(p,q)$-valued relations
$$\phi_{p,q}:X_p\oto Y_q\quad(p,q\in\CQ_{\ro}),$$
i.e., a family of maps
$$\phi_{p,q}:X_p\times Y_q\to\CQ(p,q)\quad(p,q\in\CQ_{\ro}),$$
where $\phi_{p,q}$ is the restriction of $\phi$ on the fibres $X_p$ and $Y_q$.
\end{rem}

\subsection{$\CQ$-categories}

A \emph{$\CQ$-category} (or, a \emph{category enriched in $\CQ$}) \cite{Rosenthal1996,Stubbe2005} is a $\CQ_{\ro}$-typed set $X$ equipped with a $\CQ$-relation $1_X^{\nat}:X\oto X$, such that
$$\ka_X\leq 1_X^{\nat}\quad\text{and}\quad 1_X^{\nat}\circ 1_X^{\nat}\leq 1_X^{\nat}$$
in the quantaloid $\QRel$; that is,
$$\id_{|x|}\leq 1_X^{\nat}(x,x)\quad\text{and}\quad 1_X^{\nat}(y,z)\circ 1_X^{\nat}(x,y)\leq 1_X^{\nat}(x,z)$$
for all $x,y,z\in X$. With morphisms of $\CQ$-categories given by \emph{$\CQ$-functors} $f:X\to Y$, i.e., maps $f:X\to Y$ such that
$$|x|=|fx|\quad\text{and}\quad 1_X^{\nat}(x,x')\leq 1_Y^{\nat}(fx,fx')$$
for all $x,x'\in X$, we obtain a category
$$\QCat.$$
A pair of $\CQ$-functors $f:X\to Y$, $g:Y\to X$ forms an \emph{adjunction} in $\QCat$, denoted by $f\dv g$, if
\begin{equation} \label{adj-QCat-def}
1_Y^{\nat}(fx,y)=1_X^{\nat}(x,gy)
\end{equation}
for all $x\in X$, $y\in Y$. In this case, we say that $f$ is the \emph{left adjoint} of $g$, and $g$ is the \emph{right adjoint} of $f$.

A $\CQ$-relation $\phi:X\oto Y$ between $\CQ$-categories becomes a \emph{$\CQ$-distributor} if
$$1_Y^{\nat}\circ\phi\circ 1_X^{\nat}=\phi;$$
that is,
$$1_Y^{\nat}(y,y')\circ\phi(x,y)\circ 1_X^{\nat}(x',x)\leq\phi(x',y')$$
for all $x,x'\in X$, $y,y'\in Y$. $\CQ$-categories and $\CQ$-distributors constitute a (large) quantaloid $\QDist$ in which compositions and implications are calculated as in $\QRel$; the identity $\CQ$-distributor on each $\CQ$-category $X$ is given by $1_X^{\nat}:X\oto X$.

Each $\CQ_{\ro}$-typed set $X$ is equipped with a \emph{discrete} $\CQ$-category structure, given by the identity $\CQ$-relation $\ka_X$. In particular, for each $q\in\CQ_{\ro}$, $\{q\}$ is a discrete $\CQ$-category with only one object $q$ with $|q|=q$. It is obvious that each $\CQ$-relation $\phi:X\oto Y$ can be viewed as a $\CQ$-distributor of discrete $\CQ$-categories, and thus $\QRel$ is embedded in $\QDist$ as a full subquantaloid.

A \emph{presheaf} with type $q$ on a $\CQ$-category $X$ is a $\CQ$-distributor $\mu:X\oto\{q\}$. Presheaves on $X$ constitute a $\CQ$-category $\PX$ with
$$1_{\PX}^{\nat}(\mu,\mu'):=\mu'\ldd\mu=\bw_{x\in X}\mu'(x)\ldd\mu(x)$$
for all $\mu,\mu'\in\PX$. Dually, the $\CQ$-category $\PdX$ of \emph{copresheaves} on $X$ consists of $\CQ$-distributors $\lam:\{q\}\oto X$ as objects with type $q$ $(q\in\CQ_{\ro})$, and
$$1_{\PdX}^{\nat}(\lam,\lam'):=\lam'\rdd\lam=\bw_{x\in X}\lam'(x)\rdd\lam(x)$$
for all $\lam,\lam'\in\PdX$.

A $\CQ$-category $X$ is \emph{complete} if the \emph{Yoneda embedding}
$$\sy:X\to\PX,\quad x\mapsto 1_X^{\nat}(-,x)$$
has a left adjoint in $\QCat$, given by $\sup:\PX\to X$; that is,
$$1_X^{\nat}(\sup\mu,-)=1_{\PX}^{\nat}(\mu,\sy-)=1_X^{\nat}\ldd\mu$$
for all $\mu\in\PX$. It is well known that the completeness of $X$ can also be characterized through the existence of a right adjoint of the \emph{co-Yoneda embedding} (see \cite[Proposition 5.10]{Stubbe2005})
$$\syd:X\oto\PdX,\quad x\mapsto 1_X^{\nat}(x,-),$$
given by $\inf:\PdX\to X$. It follows from \cite[Proposition 6.4]{Stubbe2005} that for any $\CQ$-category $X$, both $\PX$ and $\PdX$ are complete $\CQ$-categories.

\subsection{The underlying order of $\CQ$-categories}

Every $\CQ$-category $X$ admits a natural underlying (pre)order, given by
$$x\leq y\iff |x|=|y|=q\ \text{and}\ \id_q\leq 1_X^{\nat}(x,y)$$
for all $x,y\in X$. We write $x\cong y$ if $x\leq y$ and $y\leq x$. A $\CQ$-category $X$ is \emph{separated} if its underlying order is a partial order; that is, $x\cong y$ implies $x=y$ for all $x,y\in X$.

The underlying order of $\CQ$-categories allows us to order $\CQ$-functors as
\begin{equation} \label{QCat-local-order}
f\leq f':X\to Y\iff\forall x\in X:\ fx\leq f'x\iff\forall x\in X:\ \id_{|x|}\leq 1_Y^{\nat}(fx,f'x),
\end{equation}
and hence $\QCat$ becomes a \emph{2-category} (cf. \cite[Section XII.3]{MacLane1998}) with 2-cells given by the order \eqref{QCat-local-order}. Adjoint $\CQ$-functors defined by \eqref{adj-QCat-def} are actually internal adjunctions of the 2-category $\QCat$; that is, $f\dv g$ if, and only if,
$$1_X\leq gf\quad\text{and}\quad fg\leq 1_Y,$$
where $1_X$ and $1_Y$ are the identity $\CQ$-functors on $X$ and $Y$, respectively (cf. \cite[Lemma 2.2]{Stubbe2005}). In particular, $f$ and $g$ form a Galois connection between the underlying orders of $X$ and $Y$. More specifically, for any $q\in\CQ_{\ro}$, since the underlying order of a $\CQ$-category is defined fibrewise and $\CQ$-functors are type-preserving, the restriction
$$\bfig
\morphism/@{->}@<4pt>/<500,0>[X_q`Y_q;f]
\morphism(500,0)|b|/@{->}@<4pt>/<-500,0>[Y_q`X_q;g]
\place(240,5)[\mbox{\footnotesize{$\bot$}}]
\efig$$
of an adjunction $f\dv g$ in $\QCat$ to their $q$-fibres
$$X_q=\{x\in X\mid |x|=q\}\quad\text{and}\quad Y_q=\{y\in Y\mid |y|=q\}$$
is a Galois connection with respect to the underlying orders.

If $X$ is a separated complete $\CQ$-category, then every fibre $X_q$ of $X$ is a complete lattice with respect to its underlying order (cf. \cite[Theorem 2.8]{Shen2013a}). In particular, for any $\CQ$-category $X$, both $\PX$ and $\PdX$ are separated complete $\CQ$-categories, and thus all fibres
\begin{align*}
&(\PX)_q=\QDist(X,\{q\})=\{\mu\mid\mu:X\oto\{q\}\ \text{is a}\ \CQ\text{-distributor}\},\\
&(\PdX)_q=\QDist(\{q\},X)=\{\lam\mid\lam:\{q\}\oto X\ \text{is a}\ \CQ\text{-distributor}\}
\end{align*}
of $\PX$ and $\PdX$ are complete lattices. However, it should be cautious that the underlying order of $\PdX$ is the \emph{reverse} local order of $\QDist$; that is,
$$\lam\leq\lam'\ \text{in}\ \PdX\iff\lam'\leq\lam\ \text{in}\ \QDist.$$
In order to avoid confusion, we make the convention that the symbols $\leq$, $\vee$, $\wedge$ between $\CQ$-distributors always refer to the local order in $\QDist$ unless otherwise specified.

\begin{rem} \label{PX-fibre-discrete}
Considering a $\CQ_{\ro}$-typed set $X$ as a discrete $\CQ$-category, then $q$-fibres ($q\in\CQ_{\ro}$) of $\PX$ and $\PdX$ may be described as
\begin{align*}
&(\PX)_q=\QRel(X,\{q\})=\prod\limits_{p\in\CQ_{\ro}}\CQ(p,q)^{X_p},\\
&(\PdX)_q=\QRel(\{q\},X)=\prod\limits_{p\in\CQ_{\ro}}\CQ(q,p)^{X_p},
\end{align*}
since for each $x\in X$, $\CQ$-relations $\mu:X\oto\{q\}$ and $\lam:\{q\}\oto X$ are actually maps
$$\mu:X\to\CQ_{\arr}\quad\text{and}\quad\lam:X\to\CQ_{\arr}$$
with
$$\mu(x)\in\CQ(|x|,q)\quad\text{and}\quad\lam(x)\in\CQ(q,|x|)$$
for all $x\in X$.
\end{rem}

\section{Contexts of a multi-adjoint frame as $\CQ$-relations} \label{Multi-adjoint-frames}

Now let us formalize adjoint triples, the cornerstone of the theory of multi-adjoint concept lattices, as a special kind of quantaloids.

Recall that an \emph{adjoint triple} \cite{Medina2007,Medina2009,Medina2006} $(\with,\lda,\lua)$ with respect to posets $L_1$, $L_2$, $P$ consists of maps
$$\with:L_1\times L_2\to P,\quad\lda:P\times L_2\to L_1,\quad\lua:P\times L_1\to L_2$$
such that
\begin{equation} \label{with-imp-adj}
x\with y\leq z\iff x\leq z\lda y\iff y\leq z\lua x
\end{equation}
for all $x\in L_1$, $y\in L_2$, $z\in P$. Note that \eqref{with-imp-adj} necessarily forces
\begin{equation} \label{with-imp-order}
x\geq x',\ y\geq y',\ z\leq z'\implies x'\with y'\leq x\with y,\ z\lda y\leq z'\lda y',\ z\lua x\leq z'\lua x'
\end{equation}
for all $x,x'\in L_1$, $y,y'\in L_2$, $z,z'\in P$.

As the completeness of the posets under concern is necessary to construct \emph{concept lattices} later on, it does no harm to restrict our discussion to adjoint triples with respect to complete lattices. From now on we always assume that $L_1$, $L_2$, $P$ are complete lattices\footnote{In fact, even if $L_1$, $L_2$, $P$ are not complete, adjoint triples with respect to $L_1$, $L_2$, $P$ may be extended to their \emph{Dedekind--MacNeille completions} (see \cite[Lemma 38]{Medina2016}).}. Hence, an \emph{adjoint triple} $(\with,\lda,\lua)$ with respect to $L_1$, $L_2$, $P$ is uniquely determined by a map
$$\with:L_1\times L_2\to P$$
that preserves joins on both sides, i.e.,
$$\Big(\bv_{i\in I}x_i\Big)\with y=\bv_{i\in I}x_i\with y\quad\text{and}\quad x\with\Big(\bv_{i\in I}y_i\Big)=\bv_{i\in I}x\with y_i$$
for all $x,x_i\in L_1$, $y,y_i\in L_2$ $(i\in I)$; consequently, the maps $\lda:P\times L_2\to L_1$, $\lua:P\times L_1\to L_2$ would be uniquely determined by the Galois connections
$$\bfig
\morphism/@{->}@<4pt>/<500,0>[L_1`P;-\with y]
\morphism(500,0)|b|/@{->}@<4pt>/<-500,0>[P`L_1;-\lda y]
\place(250,5)[\mbox{\footnotesize{$\bot$}}]
\efig\quad\text{and}\quad\bfig
\morphism/@{->}@<4pt>/<500,0>[L_2`P;x\with -]
\morphism(500,0)|b|/@{->}@<4pt>/<-500,0>[P`L_2;-\lua x]
\place(250,5)[\mbox{\footnotesize{$\bot$}}]
\efig$$
induced by $\with$ for all $x\in L_1$, $y\in L_2$, which necessarily satisfy \eqref{with-imp-adj}.

It is then natural to regard $L_1$, $L_2$, $P$ as hom-sets of a quantaloid of three objects:


\begin{prop} \label{Qw-def}
Each adjoint triple $(\with,\lda,\lua)$ with respect to $L_1$, $L_2$, $P$ determines a non-trivial quantaloid $\Qw$ consisting of the following data:
\begin{itemize}
\item $(\Qw)_{\ro}=\{-1,0,1\}$;
\item $\Qw(-1,0)=L_1$, $\Qw(0,1)=L_2$, $\Qw(-1,1)=P$;
\item $\Qw(i,i)=\{\bot_{i,i},\id_i\}$ for all $i=-1,0,1$, and $\Qw(i,j)=\{\bot_{i,j}\}$ whenever $-1\leq j<i\leq 1$;
\item compositions in $\Qw$ are given by
    $$v\circ u=u\with v$$
    for all $u\in\Qw(-1,0)=L_1$, $v\in\Qw(0,1)=L_2$, and the other compositions are trivial;
\item left and right implications in $\Qw$ are given by
    $$w\ldd u=w\lua u\quad\text{and}\quad v\rdd w=w\lda v$$
    for all $u\in\Qw(-1,0)=L_1$, $v\in\Qw(0,1)=L_2$, $w\in\Qw(-1,1)=P$, and the other implications are trivial.
\end{itemize}
\end{prop}

\begin{rem} \label{Qw-obj}
Objects of the quantaloid $\Qw$ are denoted by numbers $-1$, $0$, $1$ only for the convenience of expression, so that its hom-sets $\Qw(i,j)$ can be described in a unified way. Similarly, objects of the quantaloids $\QFL$, $\QPL$, $\QOL$, respectively given by Propositions \ref{QFL-def}, \ref{QPL-def} and \ref{QOL-def} below, are denoted by numbers for the same purpose. It should be noted that the ordering of these objects is not essential for the construction of the quantaloids.
\end{rem}

The quantaloid constructed in Proposition \ref{Qw-def} can be extended to characterize the notion of \emph{multi-adjoint frame} \cite{Medina2009}. Explicitly, a \emph{multi-adjoint frame} is a tuple
$$\CL=(L_1,L_2,P,\with_1,\lda^1,\lua_1,\dots,\with_n,\lda^n,\lua_n),$$
such that $(\with_i,\lda^i,\lua_i)$ is an adjoint triple with respect to $L_1$, $L_2$, $P$ for all $i=1,\dots,n$, and it corresponds to a quantaloid of $n+2$ objects:

\begin{prop} \label{QFL-def}
Each multi-adjoint frame $\CL=(L_1,L_2,P,\with_1,\dots,\with_n)$ gives rise to a non-trivial quantaloid $\QFL$ consisting of the following data:
\begin{itemize}
\item $(\QFL)_{\ro}=\{-1,0,1,\dots,n\}$;
\item $\QFL(-1,0)=L_1$, $\QFL(0,i)=L_2$, $\QFL(-1,i)=P$ for all $i=1,\dots,n$;
\item $\QFL(i,i)=\{\bot_{i,i},\id_i\}$ for all $i=-1,0,1,\dots,n$, and $\QFL(i,j)=\{\bot_{i,j}\}$ whenever $-1\leq j<i\leq n$ or $0<i<j\leq n$;
\item compositions in $\QFL$ are given by
    $$v\circ u=u\with_i v$$
    for all $u\in\QFL(-1,0)=L_1$, $v\in\QFL(0,i)=L_2$ $(i=1,\dots,n)$, and the other compositions are trivial;
\item left and right implications in $\QFL$ are given by
    $$w\ldd u=w\lua_i u\quad\text{and}\quad v\rdd w=w\lda^i v$$
    for all $u\in\QFL(-1,0)=L_1$, $v\in\QFL(0,i)=L_2$, $w\in\QFL(-1,i)=P$ $(i=1,\dots,n)$, and the other implications are trivial.
\end{itemize}
\end{prop}

Recall that a \emph{context} \cite{Medina2009} of a multi-adjoint frame $\CL=(L_1,L_2,P,\with_1,\dots,\with_n)$ is a $P$-valued relation
$$\phi:X\oto Y,$$
i.e., a map
$$\phi:X\times Y\to P,$$
together with a map
$$|\text{-}|:Y\to\{1,\dots,n\},$$
where $X$ is interpreted as the set of \emph{properties} (also \emph{attributes}) and $Y$ the set of \emph{objects}. Therefore, contexts of a multi-adjoint frame $\CL$ are exactly relations valued in the quantaloid $\QFL$:

\begin{prop} \label{context-QFL-relation}
Let $\CL=(L_1,L_2,P,\with_1,\dots,\with_n)$ be a multi-adjoint frame and let $\QFL$ be the quantaloid determined by Proposition \ref{QFL-def}. Then a context $(X,Y,\phi)$ of $\CL$ is exactly a $\QFL$-relation $\phi_F:X\oto Y$ between $(\QFL)_{\ro}$-typed sets with
$$|x|=-1,\quad|y|\in\{1,\dots,n\}\quad\text{and}\quad\phi_F(x,y)=\phi(x,y)$$
for all $x\in X$, $y\in Y$.
\end{prop}

\section{Multi-adjoint concept lattices via Isbell adjunctions} \label{FCA-Isbell}

%

Recall that a \emph{$\CQ$-closure operator} \cite{Shen2013a} $c:X\to X$ on a $\CQ$-category $X$ is a $\CQ$-functor satisfying
$$1_X\leq c\quad\text{and}\quad cc\cong c,$$
and it follows from \cite[Propositions 3.3 and 3.5]{Shen2013a} that if $X$ is a complete $\CQ$-category, then
$$\Fix(c):=\{x\in X\mid cx\cong x\}$$
is also complete with the inherited $\CQ$-category structure from $X$. In particular, every pair of adjoint $\CQ$-functors $\bfig
\morphism/@{->}@<4pt>/<400,0>[X`Y;f]
\morphism(400,0)|b|/@{->}@<4pt>/<-400,0>[Y`X;g]
\place(200,5)[\mbox{\footnotesize{$\bot$}}]
\efig$ induces a $\CQ$-closure operator $gf:X\to X$ (see \cite[Example 3.2]{Shen2013a}).

Each $\CQ$-distributor $\phi:X\oto Y$ of $\CQ$-categories induces a pair of adjoint $\CQ$-functors
\begin{equation} \label{uphi-dphi}
\bfig
\morphism/@{->}@<4pt>/<500,0>[\PX`\PdY;\uphi]
\morphism(500,0)|b|/@{->}@<4pt>/<-500,0>[\PdY`\PX;\dphi]
\place(240,5)[\mbox{\footnotesize{$\bot$}}]
\efig
\end{equation}
in $\QCat$, called the \emph{Isbell adjunction} (see \cite[Proposition 4.1]{Shen2013a}), given by
$$\uphi\mu=\phi\ldd\mu\quad\text{and}\quad\dphi\lam=\lam\rdd\phi,$$
for all $\mu\in\PX$, $\lam\in\PdY$. In elementary words,
$$(\uphi\mu)(y)=\bw_{x\in X}\phi(x,y)\ldd\mu(x)\quad\text{and}\quad(\dphi\lam)(x)=\bw_{y\in Y}\lam(y)\rdd\phi(x,y)$$
for all $\mu\in\PX$, $y\in Y$, $\lam\in\PdY$, $x\in X$. The induced $\CQ$-closure operator $\dphi\uphi:\PX\to\PX$ generates a complete $\CQ$-category
$$\Mphi:=\Fix(\dphi\uphi)=\{\mu\in\PX\mid\dphi\uphi\mu=\mu\},$$
where ``$\cong$'' is replaced by ``$=$'' due to the separatedness of $\PX$.

\begin{rem}
Isbell adjunctions between quantaloid-enriched categories set up a very general framework of formal concept analysis (FCA).

If $\CQ={\bf 2}$ is the two-element Boolean algebra, then a ${\bf 2}$-distributor $\phi:X\oto Y$ between discrete ${\bf 2}$-categories is just a binary relation between (crisp) sets, and $\Mphi$ is the \emph{concept lattice} \cite{Davey2002,Ganter1999} of the (crisp) context $(X,Y,\phi)$.

If $\CQ$ has only one object, i.e., $\CQ=\FQ$ is a \emph{unital quantale} \cite{Rosenthal1990}, then a $\FQ$-distributor $\phi:X\oto Y$ between discrete $\FQ$-categories is a fuzzy relation between (crisp) sets (i.e., $\phi$ is a map $X\times Y\to\FQ$). Considering $(X,Y,\phi)$ as a fuzzy context of (crisp) sets $X$ and $Y$, its concept lattice is also given by $\Mphi$ (cf. \cite{Bvelohlavek2004,Lai2009,Shen2013}).

If $\CQ=\CD\FQ$ is the quantaloid of \emph{diagonals} (cf. \cite{Hoehle2011a,Pu2012,Stubbe2014}) of a quantale $\FQ$, then a $\CQ$-distributor $\phi:X\oto Y$ between discrete $\CQ$-categories is a fuzzy relation between fuzzy sets (cf. \cite[Definition 2.3]{GutierrezGarcia2018}), and the induced $\Mphi$ is the concept lattice of the fuzzy context $(X,Y,\phi)$ of fuzzy sets $X$ and $Y$ \cite{GutierrezGarcia2018,Shen2014,Shen2013b}.
\end{rem}

Now let us return to the $\QFL$-relation $\phi_F:X\oto Y$ obtained from a context $(X,Y,\phi)$ of a multi-adjoint frame $\CL=(L_1,L_2,P,\with_1,\dots,\with_n)$ in Proposition \ref{context-QFL-relation}. Considering $X$ and $Y$ as discrete $\QFL$-categories, since $|x|=-1$ and $|y|\in\{1,\dots,n\}$ for all $x\in X$, $y\in Y$, by Remark \ref{PX-fibre-discrete} we have
$$(\PX)_0=\QFL(-1,0)^X=L_1^X\quad\text{and}\quad(\PdY)_0=\prod\limits_{1\leq i\leq n}\QFL(0,i)^{Y_i}=\prod\limits_{1\leq i\leq n}L_2^{Y_i}=L_2^Y.$$
Hence, the restriction of the Isbell adjunction $(\phi_F)_{\ua}\dv(\phi_F)^{\da}$ on the $0$-fibres of $\PX$ and $\PdY$
\begin{equation} \label{uphiF-dphiF-0}
\bfig
\morphism/@{->}@<4pt>/<700,0>[(\PX)_0`(\PdY)_0;(\phi_F)_{\ua}]
\morphism(700,0)|b|/@{->}@<4pt>/<-700,0>[(\PdY)_0`(\PX)_0;(\phi_F)^{\da}]
\place(340,5)[\mbox{\footnotesize{$\bot$}}]
\efig
\end{equation}
exactly reproduces the Galois connection obtained in \cite[Proposition 7]{Medina2009}, which satisfies
\begin{align*}
&((\phi_F)_{\ua}\mu)(y)=\bw_{x\in X}\phi_F(x,y)\ldd\mu(x)=\bw_{x\in X}\phi(x,y)\lua_{|y|}\mu(x)\\
&((\phi_F)^{\da}\lam)(x)=\bw_{y\in Y}\lam(y)\rdd\phi_F(x,y)=\bw_{y\in Y}\phi(x,y)\lda^{|y|}\lam(y)
\end{align*}
for all $\mu\in(\PX)_0=L_1^X$, $y\in Y$, $\lam\in(\PdY)_0=L_2^Y$, $x\in X$.

Since the \emph{multi-adjoint concept lattice} of $(X,Y,\phi)$ is the complete lattice of fixed points of the Galois connection \eqref{uphiF-dphiF-0} (cf. \cite[Defintion 8]{Medina2009}), it is obviously given by the $0$-fibre of $\Mphi_F$:

\begin{thm} \label{FCA-lattice}
The multi-adjoint concept lattice of a context $(X,Y,\phi)$ of a multi-adjoint frame $\CL$ is isomorphic to the complete lattice $(\Mphi_F)_0$, where $\Mphi_F$ is the complete $\QFL$-category of fixed points of the Isbell adjunction \eqref{uphi-dphi} induced by the $\QFL$-relation $\phi_F:X\oto Y$ in Proposition \ref{context-QFL-relation}.
\end{thm}

\section{Multi-adjoint property-oriented and object-oriented concept lattices via Kan adjunctions} \label{RST-Kan}

Multi-adjoint object-oriented and property-oriented concept lattices introduced in \cite{Medina2012} can also be realized through adjoint functors enriched in quantaloids, and it is the goal of this section.

\subsection{Kan adjunctions}

Each $\CQ$-distributor $\phi:X\oto Y$ of $\CQ$-categories induces another two pairs of adjoint $\CQ$-functors in $\QCat$: one is the \emph{Kan adjunction} (see \cite[Proposition 5.1]{Shen2013a})
\begin{equation} \label{phi-star}
\bfig
\morphism/@{->}@<4pt>/<500,0>[\PY`\PX;\phi^*]
\morphism(500,0)|b|/@{->}@<4pt>/<-500,0>[\PX`\PY;\phi_*]
\place(240,5)[\mbox{\footnotesize{$\bot$}}]
\efig
\end{equation}
given by
$$\phi^*\lam=\lam\circ\phi\quad\text{and}\quad\phi_*\mu=\mu\ldd\phi,$$
which are calculated as
$$(\phi^*\lam)(x)=\bv_{y\in Y}\lam(y)\circ\phi(x,y)\quad\text{and}\quad(\phi_*\mu)(y)=\bw_{x\in X}\mu(x)\ldd\phi(x,y)$$
for all $\lam\in\PY$, $x\in X$, $\mu\in\PX$, $y\in Y$; the other is the \emph{dual Kan adjunction} (see \cite[Proposition 6.2.1]{Shen2014})
\begin{equation} \label{phi-dag}
\bfig
\morphism/@{->}@<4pt>/<500,0>[\PdY`\PdX;\phi_{\dag}]
\morphism(500,0)|b|/@{->}@<4pt>/<-500,0>[\PdX`\PdY;\phi^{\dag}]
\place(240,5)[\mbox{\footnotesize{$\bot$}}]
\efig
\end{equation}
given by
$$\phi_{\dag}\lam=\phi\rdd\lam\quad\text{and}\quad\phi^{\dag}\mu=\phi\circ\mu,$$
which are calculated as
$$(\phi_{\dag}\lam)(x)=\bw_{y\in Y}\phi(x,y)\rdd\lam(y)\quad\text{and}\quad(\phi^{\dag}\mu)(y)=\bv_{x\in X}\phi(x,y)\circ\mu(x)$$
for all $\lam\in\PdY$, $x\in X$, $\mu\in\PdX$, $y\in Y$. The induced $\CQ$-closure operators $\phi_*\phi^*:\PY\to\PY$ and $\phi^{\dag}\phi_{\dag}:\PdY\to\PdY$ give rise to complete $\CQ$-categories
$$\Kphi:=\Fix(\phi_*\phi^*)=\{\lam\in\PY\mid\phi_*\phi^*\lam=\lam\}\quad\text{and}\quad\Kdphi:=\Fix(\phi^{\dag}\phi_{\dag})=\{\lam\in\PdY\mid\phi^{\dag}\phi_{\dag}\lam=\lam\}.$$


\begin{rem}
The complete $\CQ$-categories $\Kphi$ and $\Kdphi$ present a categorical extension of concept lattices based on rough set theory (RST). In the case of $\CQ={\bf 2}$, considering $X$ as the (discrete) set of properties and $Y$ as the (discrete) set of objects, $\Kphi$ and $\Kdphi$ are respectively the \emph{property-oriented concept lattice} and the \emph{object-oriented concept lattice} of the (crisp) context $(X,Y,\phi)$ introduced in \cite{Yao2004a,Yao2004}, which have also been generalized to those of fuzzy contexts of (crisp) sets \cite{Georgescu2004,Lai2009,Popescu2004,Shen2013} and fuzzy contexts of fuzzy sets \cite{GutierrezGarcia2018,Shen2014}.
\end{rem}

\subsection{Multi-adjoint property-oriented concept lattices as fixed points of Kan adjunctions}

Recall that a \emph{multi-adjoint property-oriented frame} \cite{Medina2012} is a tuple
$$\CL=(L_1,L_2,P,\with_1,\lda^1,\lua_1,\dots,\with_n,\lda^n,\lua_n),$$
such that $(\with_i,\lda^i,\lua_i)$ is an adjoint triple with respect to $P$, $L_2$, $L_1$ for all $i=1,\dots,n$; that is, the maps
$$\with_i:P\times L_2\to L_1,\quad\lda^i:L_1\times L_2\to P,\quad\lua_i:L_1\times P\to L_2$$
satisfy
$$z\with_i y\leq x\iff z\leq x\lda^i y\iff y\leq x\lua_i z$$
for all $z\in P$, $y\in L_2$, $x\in L_1$. With a suitable modification of Proposition \ref{QFL-def} we may construct a quantaloid $\QPL$ from a multi-adjoint property-oriented frame $\CL$:

\begin{prop} \label{QPL-def}
Each multi-adjoint property-oriented frame $\CL=(L_1,L_2,P,\with_1,\dots,\with_n)$ gives rise to a non-trivial quantaloid $\QPL$ consisting of the following data:
\begin{itemize}
\item $(\QPL)_{\ro}=\{0,1,\dots,n,\infty\}$;
\item $\QPL(0,i)=P$, $\QPL(i,\infty)=L_2$, $\QPL(0,\infty)=L_1$ for all $i=1,\dots,n$;
\item $\QPL(i,i)=\{\bot_{i,i},\id_i\}$ for all $i=0,1,\dots,n,\infty$, and $\QPL(i,j)=\{\bot_{i,j}\}$ whenever $0\leq j<i\leq\infty$ or $0<i<j<\infty$;
\item compositions in $\QPL$ are given by
    $$v\circ u=u\with_i v$$
    for all $u\in\QPL(0,i)=P$, $v\in\QPL(i,\infty)=L_2$ $(i=1,\dots,n)$, and the other compositions are trivial;
\item left and right implications in $\QPL$ are given by
    $$w\ldd u=w\lua_i u\quad\text{and}\quad v\rdd w=w\lda^i v$$
    for all $u\in\QPL(0,i)=P$, $v\in\QPL(i,\infty)=L_2$, $w\in\QPL(0,\infty)=L_1$ $(i=1,\dots,n)$, and the other implications are trivial.
\end{itemize}
\end{prop}

A \emph{context} \cite{Medina2012} of a multi-adjoint property-oriented frame $\CL=(L_1,L_2,P,\with_1,\dots,\with_n)$ is also defined as a $P$-valued relation
$$\phi:X\oto Y$$
equipped with a map
$$|\text{-}|:Y\to\{1,\dots,n\},$$
where $X$ is interpreted as the set of \emph{properties} and $Y$ the set of \emph{objects}. Therefore:

\begin{prop} \label{context-QPL-relation}
Let $\CL=(L_1,L_2,P,\with_1,\dots,\with_n)$ be a multi-adjoint property-oriented frame and let $\QPL$ be the quantaloid determined by Proposition \ref{QPL-def}. Then a context $(X,Y,\phi)$ of $\CL$ is exactly a $\QPL$-relation $\phi_P:X\oto Y$ between $(\QPL)_{\ro}$-typed sets with
$$|x|=0,\quad |y|\in\{1,\dots,n\}\quad\text{and}\quad\phi_P(x,y)=\phi(x,y)$$
for all $x\in X$, $y\in Y$.
\end{prop}

Considering the $\QPL$-relation $\phi_P:X\oto Y$ obtained in Proposition \ref{context-QPL-relation}, by Remark \ref{PX-fibre-discrete} we have
$$(\PX)_{\infty}=\QPL(0,\infty)^X=L_1^X\quad\text{and}\quad(\PY)_{\infty}=\prod\limits_{1\leq i\leq n}\QPL(i,\infty)^{Y_i}=\prod\limits_{1\leq i\leq n}L_2^{Y_i}=L_2^Y.$$
Hence, by restricting the Kan adjunction $(\phi_P)^*\dv(\phi_P)_*$ on the $\infty$-fibres of $\PY$ and $\PX$
\begin{equation} \label{phiP-star-0}
\bfig
\morphism/@{->}@<4pt>/<700,0>[(\PY)_{\infty}`(\PX)_{\infty};(\phi_P)^*]
\morphism(700,0)|b|/@{->}@<4pt>/<-700,0>[(\PX)_{\infty}`(\PY)_{\infty};(\phi_P)_*]
\place(340,5)[\mbox{\footnotesize{$\bot$}}]
\efig
\end{equation}
we obtain the Galois connection given in \cite[Section 4]{Medina2012}, which satisfies
\begin{align*}
&((\phi_P)^*\lam)(x)=\bv_{y\in Y}\lam(y)\circ\phi_P(x,y)=\bv_{y\in Y}\phi(x,y)\with_{|y|}\lam(y)\\
&((\phi_P)_*\mu)(y)=\bw_{x\in X}\mu(x)\ldd\phi_P(x,y)=\bw_{x\in X}\mu(x)\lua_{|y|}\phi(x,y)
\end{align*}
for all $\lam\in(\PY)_{\infty}=L_2^Y$, $x\in X$, $\mu\in(\PX)_{\infty}=L_1^X$, $y\in Y$.

Since the \emph{multi-adjoint property-oriented concept lattice} of $(X,Y,\phi)$ is the complete lattice of fixed points of the Galois connection \eqref{phiP-star-0} (cf. \cite[Section 4]{Medina2012}), it is obviously given by the $\infty$-fibre of $\Kphi_P$:

\begin{thm} \label{RST-lattice-P}
The multi-adjoint property-oriented concept lattice of a context $(X,Y,\phi)$ of a multi-adjoint property-oriented frame $\CL$ is isomorphic to the complete lattice $(\Kphi_P)_{\infty}$, where $\Kphi_P$ is the complete $\QPL$-category of fixed points of the Kan adjunction \eqref{phi-star} induced by the $\QPL$-relation $\phi_P:X\oto Y$ in Proposition \ref{context-QPL-relation}.
\end{thm}

\subsection{Multi-adjoint object-oriented concept lattices as fixed points of dual Kan adjunctions}

Following the terminology of \cite[Section 5]{Medina2012}, a \emph{multi-adjoint object-oriented frame} is a tuple
$$\CL=(L_1,L_2,P,\with_1,\lda^1,\lua_1,\dots,\with_n,\lda^n,\lua_n),$$
such that $(\with_i,\lda^i,\lua_i)$ is an adjoint triple with respect to $L_1$, $P$, $L_2$ for all $i=1,\dots,n$; that is, the maps
$$\with_i:L_1\times P\to L_2,\quad\lda^i:L_2\times P\to L_1,\quad\lua_i:L_2\times L_1\to P$$
satisfy
$$x\with_i z\leq y\iff x\leq y\lda^i z\iff z\leq y\lua_i x$$
for all $x\in L_1$, $z\in P$, $y\in L_2$. Similarly as in Proposition \ref{QPL-def} we may construct a quantaloid $\QOL$:

\begin{prop} \label{QOL-def}
Each multi-adjoint object-oriented frame $\CL=(L_1,L_2,P,\with_1,\dots,\with_n)$ gives rise to a non-trivial quantaloid $\QOL$ consisting of the following data:
\begin{itemize}
\item $(\QOL)_{\ro}=\{-1,0,1,\dots,n\}$;
\item $\QOL(-1,0)=L_1$, $\QOL(0,i)=P$, $\QOL(-1,i)=L_2$ for all $i=1,\dots,n$;
\item $\QOL(i,i)=\{\bot_{i,i},\id_i\}$ for all $i=-1,0,1,\dots,n$, and $\QOL(i,j)=\{\bot_{i,j}\}$ whenever $-1\leq j<i\leq n$ or $0<i<j\leq n$;
\item compositions in $\QOL$ are given by
    $$v\circ u=u\with_i v$$
    for all $u\in\QOL(-1,0)=L_1$, $v\in\QOL(0,i)=P$ $(i=1,\dots,n)$, and the other compositions are trivial;
\item left and right implications in $\QOL$ are given by
    $$w\ldd u=w\lua_i u\quad\text{and}\quad v\rdd w=w\lda^i v$$
    for all $u\in\QOL(-1,0)=L_1$, $v\in\QOL(0,i)=P$, $w\in\QOL(-1,i)=L_2$ $(i=1,\dots,n)$, and the other implications are trivial.
\end{itemize}
\end{prop}

With a \emph{context} \cite{Medina2012} of a multi-adjoint object-oriented frame $\CL=(L_1,L_2,P,\with_1,\dots,\with_n)$ defined as a $P$-valued relation
$$\phi:X\oto Y$$
equipped with a map
$$|\text{-}|:Y\to\{1,\dots,n\},$$
where elements in $X$ and $Y$ are \emph{properties} and \emph{objects}, respectively, we deduce the following parallel proposition of \ref{context-QPL-relation}:

\begin{prop} \label{context-QOL-relation}
Let $\CL=(L_1,L_2,P,\with_1,\dots,\with_n)$ be a multi-adjoint object-oriented frame and let $\QOL$ be the quantaloid determined by Proposition \ref{QOL-def}. Then a context $(X,Y,\phi)$ of $\CL$ is exactly a $\QOL$-relation $\phi_O:X\oto Y$ between $(\QOL)_{\ro}$-typed sets with
$$|x|=0,\quad|y|\in\{1,\dots,n\}\quad\text{and}\quad\phi_O(x,y)=\phi(x,y)$$
for all $x\in X$, $y\in Y$.
\end{prop}

For the above $\QOL$-relation $\phi_O:X\oto Y$, with Remark \ref{PX-fibre-discrete} it is easy to see that
$$(\PdX)_{-1}=\QOL(-1,0)^X=L_1^X\quad\text{and}\quad(\PdY)_{-1}=\prod\limits_{1\leq i\leq n}\QOL(-1,i)^{Y_i}=\prod\limits_{1\leq i\leq n}L_2^{Y_i}=L_2^Y.$$
Consequently, by restricting the dual Kan adjunction $(\phi_O)_{\dag}\dv(\phi_O)^{\dag}$ on the $(-1)$-fibres of $\PdY$ and $\PdX$
\begin{equation} \label{phiO-dag-0}
\bfig
\morphism/@{->}@<4pt>/<800,0>[(\PdY)_{-1}`(\PdX)_{-1};(\phi_O)_{\dag}]
\morphism(800,0)|b|/@{->}@<4pt>/<-800,0>[(\PdX)_{-1}`(\PdY)_{-1};(\phi_O)^{\dag}]
\place(390,5)[\mbox{\footnotesize{$\bot$}}]
\efig
\end{equation}
we obtain the Galois connection given in \cite[Section 5]{Medina2012}, which satisfies
\begin{align*}
&((\phi_O)_{\dag}\lam)(x)=\bw_{y\in Y}\phi_O(x,y)\rdd\lam(y)=\bw_{y\in Y}\lam(y)\lda^{|y|}\phi(x,y)\\
&((\phi_O)^{\dag}\mu)(y)=\bv_{x\in X}\phi_O(x,y)\circ\mu(x)=\bv_{x\in X}\mu(x)\with_{|y|}\phi(x,y)
\end{align*}
for all $\lam\in(\PdY)_{-1}=L_2^Y$, $x\in X$, $\mu\in(\PdX)_{-1}=L_1^X$, $y\in Y$.

As the \emph{multi-adjoint object-oriented concept lattice} of $(X,Y,\phi)$ is the complete lattice of fixed points of the Galois connection \eqref{phiO-dag-0} (cf. \cite[Section 5]{Medina2012}), it is clearly given by the $(-1)$-fibre of $\Kdphi_O$:

\begin{thm} \label{RST-lattice-O}
The multi-adjoint object-oriented concept lattice of a context $(X,Y,\phi)$ of a multi-adjoint object-oriented frame $\CL$ is isomorphic to the complete lattice $(\Kdphi_O)_{-1}$, where $\Kdphi_O$ is the complete $\QOL$-category of fixed points of the dual Kan adjunction \eqref{phi-dag} induced by the $\QOL$-relation $\phi_O:X\oto Y$ in Proposition \ref{context-QOL-relation}.
\end{thm}


\section{Concluding remarks}

In category theory we are interested not only in \emph{categories of structures}, but also in \emph{categories as structures}; concept lattices in the theories of FCA and RST are typical instances of the latter. Considering a distributor between quantaloid-enriched categories as a multi-typed and multi-valued relation, our recent work \cite{Lai2017} extends the machinery of FCA and RST to the general framework of quantaloid-enriched categories.

The main results of this paper, Theorems \ref{FCA-lattice}, \ref{RST-lattice-P} and \ref{RST-lattice-O}, reveal that multi-adjoint concept lattices, multi-adjoint property-oriented concept lattices and multi-adjoint object-oriented concept lattices are also instances of quantaloid-enriched categories, which justify again the importance of the quantaloidal approach in the study of FCA and RST.

We end this paper with two questions to be considered in future works:
\begin{enumerate}[label=(\arabic*)]
\item Can we apply the representation theorems obtained in \cite{Lai2017} to derive more representation theorems of multi-adjoint (property/object-oriented) concept lattices?
\item As Isbell adjunctions and Kan adjunctions make sense not only for quantaloid-enriched categories, but also for general (enriched) categories, is it possible to establish the theories of FCA and RST in the framework of general (enriched) categories?
\end{enumerate}

\section*{Acknowledgement}

The authors acknowledge the support of National Natural Science Foundation of China (No. 11771310 and No. 11701396). We thank Professor Dexue Zhang for bringing the topic of this paper to our attention, and we thank the anonymous referees for several helpful remarks.





\begin{thebibliography}{10}

\bibitem{Bvelohlavek2004}
R.~B{\v e}lohl{\' a}vek.
\newblock Concept lattices and order in fuzzy logic.
\newblock {\em Annals of Pure and Applied Logic}, 128(1-3):277--298, 2004.

\bibitem{Cornejo2015}
M.~E. Cornejo, J.~Medina, and E.~Ram{\' \i}rez-Poussa.
\newblock Attribute reduction in multi-adjoint concept lattices.
\newblock {\em Information Sciences}, 294:41--56, 2015.

\bibitem{Cornejo2015a}
M.~E. Cornejo, J.~Medina, and E.~Ram{\' \i}rez-Poussa.
\newblock On the use of irreducible elements for reducing multi-adjoint concept
  lattices.
\newblock {\em Knowledge-Based Systems}, 89:192--202, 2015.

\bibitem{Cornejo2018}
M.~E. Cornejo, J.~Medina, and E.~Ram{\' \i}rez-Poussa.
\newblock Characterizing reducts in multi-adjoint concept lattices.
\newblock {\em Information Sciences}, 422:364--376, 2018.

\bibitem{Cornelis2014}
C.~Cornelis, J.~Medina, and N.~Verbiest.
\newblock Multi-adjoint fuzzy rough sets: Definition, properties and attribute
  selection.
\newblock {\em International Journal of Approximate Reasoning}, 55(1):412--426,
  2014.

\bibitem{Davey2002}
B.~A. Davey and H.~A. Priestley.
\newblock {\em Introduction to Lattices and Order}.
\newblock Cambridge University Press, Cambridge, second edition, 2002.

\bibitem{Ganter1999}
B.~Ganter and R.~Wille.
\newblock {\em Formal Concept Analysis: Mathematical Foundations}.
\newblock Springer, Berlin--Heidelberg, 1999.

\bibitem{Georgescu2004}
G.~Georgescu and A.~Popescu.
\newblock Non-dual fuzzy connections.
\newblock {\em Archive for Mathematical Logic}, 43:1009--1039, 2004.

\bibitem{GutierrezGarcia2018}
J.~Guti\'{e}rrez~Garc\'{\i}a, H.~Lai, and L.~Shen.
\newblock Fuzzy {Galois} connections on fuzzy sets.
\newblock {\em Fuzzy Sets and Systems}, 352:26--55, 2018.

\bibitem{Heymans2010}
H.~Heymans.
\newblock {\em Sheaves on Quantales as Generalized Metric Spaces}.
\newblock PhD thesis, Universiteit Antwerpen, Belgium, 2010.

\bibitem{Hoehle2015}
U.~H{\"o}hle.
\newblock Many-valued preorders {I}: The basis of many-valued mathematics.
\newblock In L.~Magdalena, J.~L. Verdegay, and F.~Esteva, editors, {\em Enric
  Trillas: A Passion for Fuzzy Sets: A Collection of Recent Works on Fuzzy
  Logic}, volume 322 of {\em Studies in Fuzziness and Soft Computing}, pages
  125--150. Springer, Cham, 2015.

\bibitem{Hoehle2011a}
U.~H{\"o}hle and T.~Kubiak.
\newblock A non-commutative and non-idempotent theory of quantale sets.
\newblock {\em Fuzzy Sets and Systems}, 166:1--43, 2011.

\bibitem{Lai2017}
H.~Lai and L.~Shen.
\newblock Fixed points of adjoint functors enriched in a quantaloid.
\newblock {\em Fuzzy Sets and Systems}, 321:1--28, 2017.

\bibitem{Lai2009}
H.~Lai and D.~Zhang.
\newblock Concept lattices of fuzzy contexts: Formal concept analysis vs. rough
  set theory.
\newblock {\em International Journal of Approximate Reasoning}, 50(5):695--707,
  2009.

\bibitem{Lawvere1973}
F.~W. Lawvere.
\newblock Metric spaces, generalized logic and closed categories.
\newblock {\em Rendiconti del Seminario Mat\'{e}matico e Fisico di Milano},
  XLIII:135--166, 1973.

\bibitem{MacLane1998}
S.~Mac~Lane.
\newblock {\em Categories for the Working Mathematician}, volume~5 of {\em
  Graduate Texts in Mathematics}.
\newblock Springer, New York, second edition, 1998.

\bibitem{Medina2012}
J.~Medina.
\newblock Multi-adjoint property-oriented and object-oriented concept lattices.
\newblock {\em Information Sciences}, 190:95--106, 2012.

\bibitem{Medina2012a}
J.~Medina and M.~Ojeda-Aciego.
\newblock On multi-adjoint concept lattices based on heterogeneous conjunctors.
\newblock {\em Fuzzy Sets and Systems}, 208:95--110, 2012.

\bibitem{Medina2016}
J.~Medina, M.~Ojeda-Aciego, J.~P{\' o}cs, and E.~Ram{\' \i}rez-Poussa.
\newblock On the {Dedekind--MacNeille} completion and formal concept analysis
  based on multilattices.
\newblock {\em Fuzzy Sets and Systems}, 303:1--20, 2016.

\bibitem{Medina2007}
J.~Medina, M.~Ojeda-Aciego, and J.~Ruiz-Calvi{\~{n}}o.
\newblock On multi-adjoint concept lattices: Definition and representation
  theorem.
\newblock In S.~O. Kuznetsov and S.~Schmidt, editors, {\em Formal Concept
  Analysis: 5th International Conference, ICFCA 2007}, volume 4390 of {\em
  Lecture Notes in Computer Science}, pages 197--209. Springer,
  Berlin--Heidelberg, 2007.

\bibitem{Medina2009}
J.~Medina, M.~Ojeda-Aciego, and J.~Ruiz-Calvi{\~ n}o.
\newblock Formal concept analysis via multi-adjoint concept lattices.
\newblock {\em Fuzzy Sets and Systems}, 160(2):130--144, 2009.

\bibitem{Medina2006}
J.~Medina and J.~Ruiz-Calvi{\~ n}o.
\newblock Towards multi-adjoint concept lattices.
\newblock In {\em Information Processing and Management of Uncertainty for
  Knowledge-Based Systems, IPMU}, pages 2566--2571, 2006.

\bibitem{Pawlak1982}
Z.~Pawlak.
\newblock Rough sets.
\newblock {\em International Journal of Computer $\&$ Information Sciences},
  11(5):341--356, 1982.

\bibitem{Polkowski2002}
L.~Polkowski.
\newblock {\em Rough Sets: Mathematical Foundations}, volume~15 of {\em
  Advances in Intelligent and Soft Computing}.
\newblock Physica-Verlag, Heidelberg, 2002.

\bibitem{Popescu2004}
A.~Popescu.
\newblock A general approach to fuzzy concepts.
\newblock {\em Mathematical Logic Quarterly}, 50(3):265--280, 2004.

\bibitem{Pu2012}
Q.~Pu and D.~Zhang.
\newblock Preordered sets valued in a {GL}-monoid.
\newblock {\em Fuzzy Sets and Systems}, 187(1):1--32, 2012.

\bibitem{Rosenthal1990}
K.~I. Rosenthal.
\newblock {\em Quantales and their Applications}, volume 234 of {\em Pitman
  research notes in mathematics series}.
\newblock Longman, Harlow, 1990.

\bibitem{Rosenthal1996}
K.~I. Rosenthal.
\newblock {\em The Theory of Quantaloids}, volume 348 of {\em Pitman Research
  Notes in Mathematics Series}.
\newblock Longman, Harlow, 1996.

\bibitem{Shen2014}
L.~Shen.
\newblock {\em Adjunctions in Quantaloid-enriched Categories}.
\newblock PhD thesis, Sichuan University, Chengdu, 2014.

\bibitem{Shen2013a}
L.~Shen and D.~Zhang.
\newblock Categories enriched over a quantaloid: {Isbell} adjunctions and {Kan}
  adjunctions.
\newblock {\em Theory and Applications of Categories}, 28(20):577--615, 2013.

\bibitem{Shen2013}
L.~Shen and D.~Zhang.
\newblock The concept lattice functors.
\newblock {\em International Journal of Approximate Reasoning}, 54(1):166--183,
  2013.

\bibitem{Shen2013b}
L.~Shen and D.~Zhang.
\newblock Formal concept analysis on fuzzy sets.
\newblock In {\em Proceedings of the 2013 Joint IFSA World Congress and NAFIPS
  Annual Meeting (IFSA/NAFIPS)}, pages 215--219. IEEE, 2013.

\bibitem{Stubbe2005}
I.~Stubbe.
\newblock Categorical structures enriched in a quantaloid: categories,
  distributors and functors.
\newblock {\em Theory and Applications of Categories}, 14(1):1--45, 2005.

\bibitem{Stubbe2006}
I.~Stubbe.
\newblock Categorical structures enriched in a quantaloid: tensored and
  cotensored categories.
\newblock {\em Theory and Applications of Categories}, 16(14):283--306, 2006.

\bibitem{Stubbe2014}
I.~Stubbe.
\newblock An introduction to quantaloid-enriched categories.
\newblock {\em Fuzzy Sets and Systems}, 256:95--116, 2014.

\bibitem{Walters1981}
R.~F.~C. Walters.
\newblock Sheaves and {Cauchy}-complete categories.
\newblock {\em Cahiers de Topologie et G{\'e}om{\'e}trie Diff{\'e}rentielle
  Cat{\'e}goriques}, 22(3):283--286, 1981.

\bibitem{Yao2004a}
Y.~Yao.
\newblock A comparative study of formal concept analysis and rough set theory
  in data analysis.
\newblock In S.~Tsumoto, R.~S{\l}owi{\'{n}}ski, J.~Komorowski, and J.~W.
  Grzyma{\l}a-Busse, editors, {\em Rough Sets and Current Trends in Computing},
  pages 59--68. Springer, 2004.

\bibitem{Yao2004}
Y.~Yao.
\newblock Concept lattices in rough set theory.
\newblock In {\em Proceedings of 2004 Annual Meeting of the North American
  Fuzzy Information Processing Society (NAFIPS 2004)}, volume~2, pages
  796--801. IEEE, 2004.

\end{thebibliography}

\end{document}